\documentclass{ws-p8-50x6-00}
\begin{document}

\title{Coherent Photon-Pomeron and Photon-Photon Interactions
in Ultra-Peripheral Collisions at RHIC}
\author{F. Meissner}
\address{ for the STAR Collaboration \\Lawrence Berkeley National Laboratory\\ 
70-319,  Berkeley, CA, 94720 USA \\ E-mail: FMeissner@lbl.gov}
\maketitle

%\classification{43.35.Ei, 78.60.Mq}
%\keywords{heavy ion collisions, vector meson production}

\abstracts{ Ultra-peripheral heavy ion  collisions involve long range electromagnetic
interactions at impact parameters larger than twice the nuclear radius,
where no nucleon-nucleon collisions occur. 
The first observation of coherent $\rho^0$ production  with and without accompanying nuclear 
breakup, $AuAu\! \rightarrow\! Au^\star Au^\star \rho^0$ and  $AuAu\! \rightarrow Au Au\!
 \rho^0$ respectively, and the observation of $e^+e^-$ pair production  
$AuAu \!\rightarrow \!Au^\star Au^\star e^+ e^-$ are presented by the STAR collaboration. 
The  transverse momentum spectra are  peaked at low $p_T$,
showing the coherent coupling to the nuclei. 
A  clear $\rho^0$ signal is observed in the two pion invariant  mass spectrum.}

%%%% \section{Introduction}
\begin{figure}[!b]
\vspace*{-.7cm}
\begin{center}
\leavevmode
\epsfclipon
\epsfysize=3.5  cm
\epsfbox{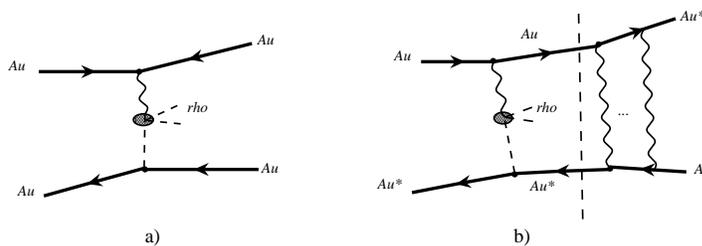}
\end{center}
\vskip -.25 in
\caption[]{Feynman diagrams for (a) $\rho^0$ production, and (b)
$\rho^0$ production with nuclear excitation.
Factorization is indicated by the dashed line; nuclear excitation 
may occur by the exchange of a single or multiple photons. }
\label{fig:feynman}
\end{figure}
In ultra-peripheral heavy ion collisions, photo-nuclear interactions
take place at impact parameters $b$ larger than twice the nuclear radius $R_A$,
where no nucleon-nucleon collisions occur~\cite{baurrev}. 
The large charge of a relativistic heavy nucleus is a 
strong source of quasi-real photons. Exclusive $\rho^0$ meson production
$AuAu\! \rightarrow\! Au Au \rho^0$ can then  be described by the  vector meson 
dominance model~\cite{sakurai}:
a photon emitted by one nucleus  fluctuates 
to a virtual quark-anti quark  pair; this intermediate  state scatters diffractively 
from the other nucleus, emerging  as a vector meson. 
The diagram for this process is shown in Fig.~\ref{fig:feynman}a. 
Here, the gold nuclei remain  in their ground state.
Additional photon exchange can yield nuclear excitation and the subsequent
emission of single or multiple neutrons, as shown in Fig.~\ref{fig:feynman}b 
for the process  $AuAu \rightarrow Au^\star Au^\star \rho^0$.

Photon and Pomeron can couple coherently to the spatially extended
electric and nuclear charge of the gold nuclei. A coherence condition
follows from  the uncertainty principle: a low transverse momentum of 
$p_T \!<\! 2 \hbar/ R_A$ ($\!\sim\!100$~MeV at RHIC for $R_A \sim 7$fm),
and a maximum  longitudinal momentum  of $p_\| \!< \!2 \hbar \gamma / R_A$ ($\!\sim\!6$~GeV at RHIC).
The coupling strength of the photon is proportional to  the square of the 
charge $Z^2$ (6241 for $Au$);  the strength of the Pomeron coupling lies 
between $A^{4/3}$ for surface coupling to $A^2$ in the  bulk limit ($10^3$ to $10^4$ for $Au$). 
This leads to large cross sections; for gold collisions at
$\sqrt{s_{NN}}\!=\!130$~GeV the $\rho^0$ production cross section is
expected to be about 400 mb or 5$\%$ of the total hadronic cross section\cite{vmrates}.

It is impossible to determine which nucleus was the photon source or the target, thus 
the amplitudes for $\rho^0$ production from both ions interfere.
Since the $\rho^0$ has negative parity, this interference is  destructive.
The short-lived $\rho^0$ decay before they travel the distance of the impact parameter $b$, 
and the interference is believed to be  sensitive to the post-decay wave function\cite{interfere}. 

%% STAR ==
\vspace*{-.4cm}
\begin{figure}[!h]
\begin{minipage}{0.55\textwidth}
\hspace*{0.1cm} In the year 2000, RHIC collided 
gold nuclei at a center-of-mass energy of $\sqrt{s_{NN}}\!=\! 130$~GeV/nucleon. 
The STAR detector consists of a 4.2~m long cylindrical time projection chamber (TPC) of  2~m radius. In 2000 the TPC was operated
in a 0.25~T  solenoidal magnetic field.
Particles were  identified by their energy loss  in the TPC.
A central trigger barrel (CTB) of scintillators surrounds the TPC. 
Two zero degree calorimeters (ZDC) at $\pm$ 18m  from the interaction point are sensitive to 
the neutral  remnants of  nuclear breakup.
\end{minipage}
\hspace*{0.05\textwidth}
\begin{minipage}{0.4\textwidth}
{\includegraphics[height=.2\textheight,angle=180,clip=true]{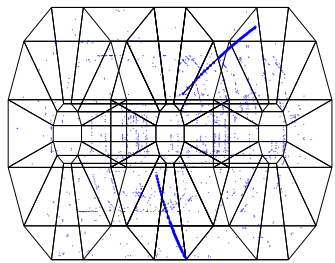}}
\caption{Display of a  typical $\rho^0$ candidate event in the STAR TPC. \label{fig:event}}
\end{minipage}
\end{figure}

\vspace*{-0.5cm}
\noindent
Ultra-peripheral collisions  have a  specific  experimental signature:
the $\pi^+\pi^-$ decay products of the $\rho^0$  meson are observed
in an otherwise 'empty'  spectrometer.  Fig.~\ref{fig:event} shows a  typical event 
candidate; the tracks are approximately  back-to-back in the transverse plane
due to the  small $p_T$ of the pair.

\begin{figure}[!h]
\resizebox{.45\textwidth}{!}{\includegraphics[height=.5\textheight]{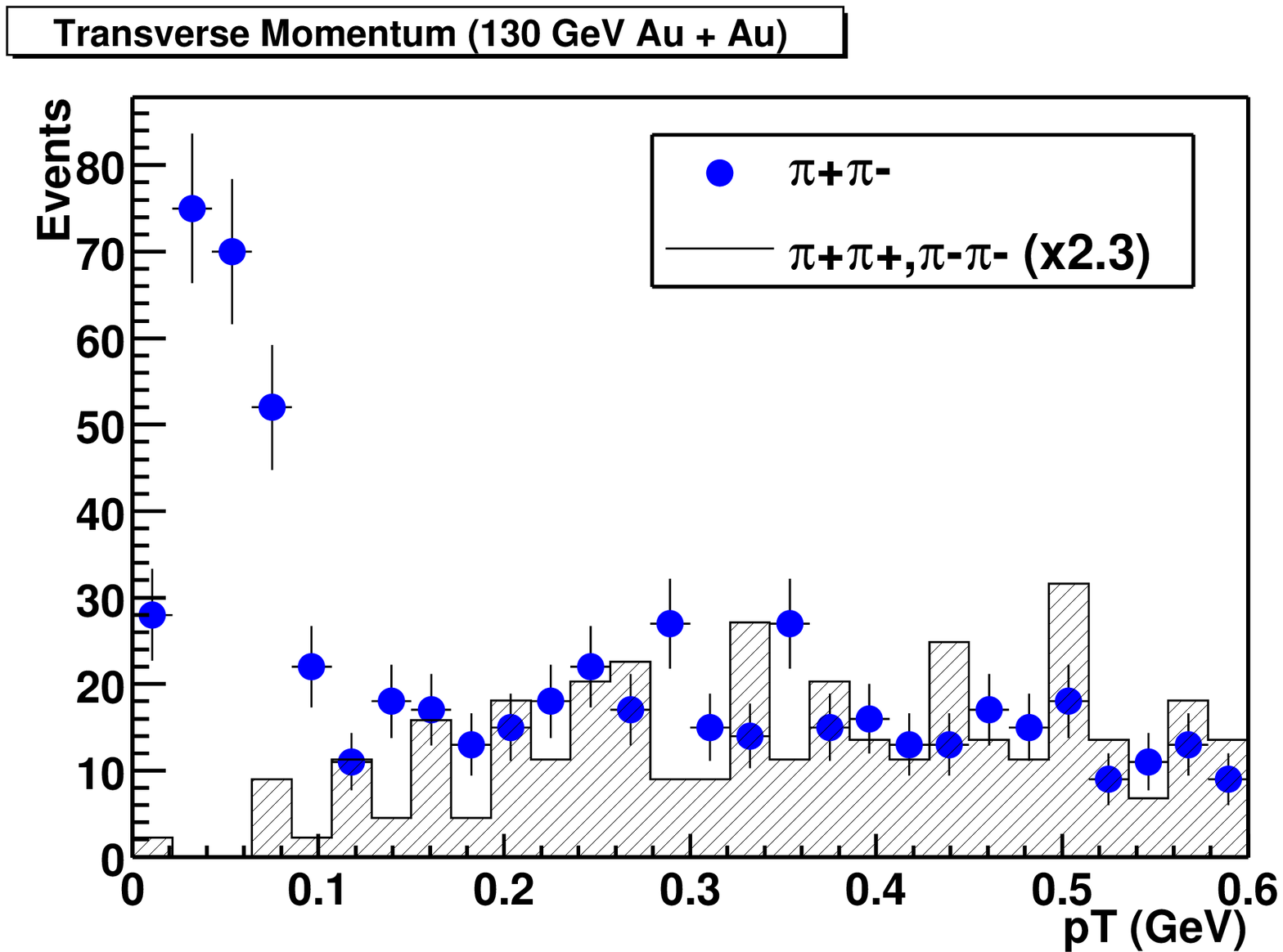}}
\resizebox{.45\textwidth}{!}{\includegraphics[height=.5\textheight]{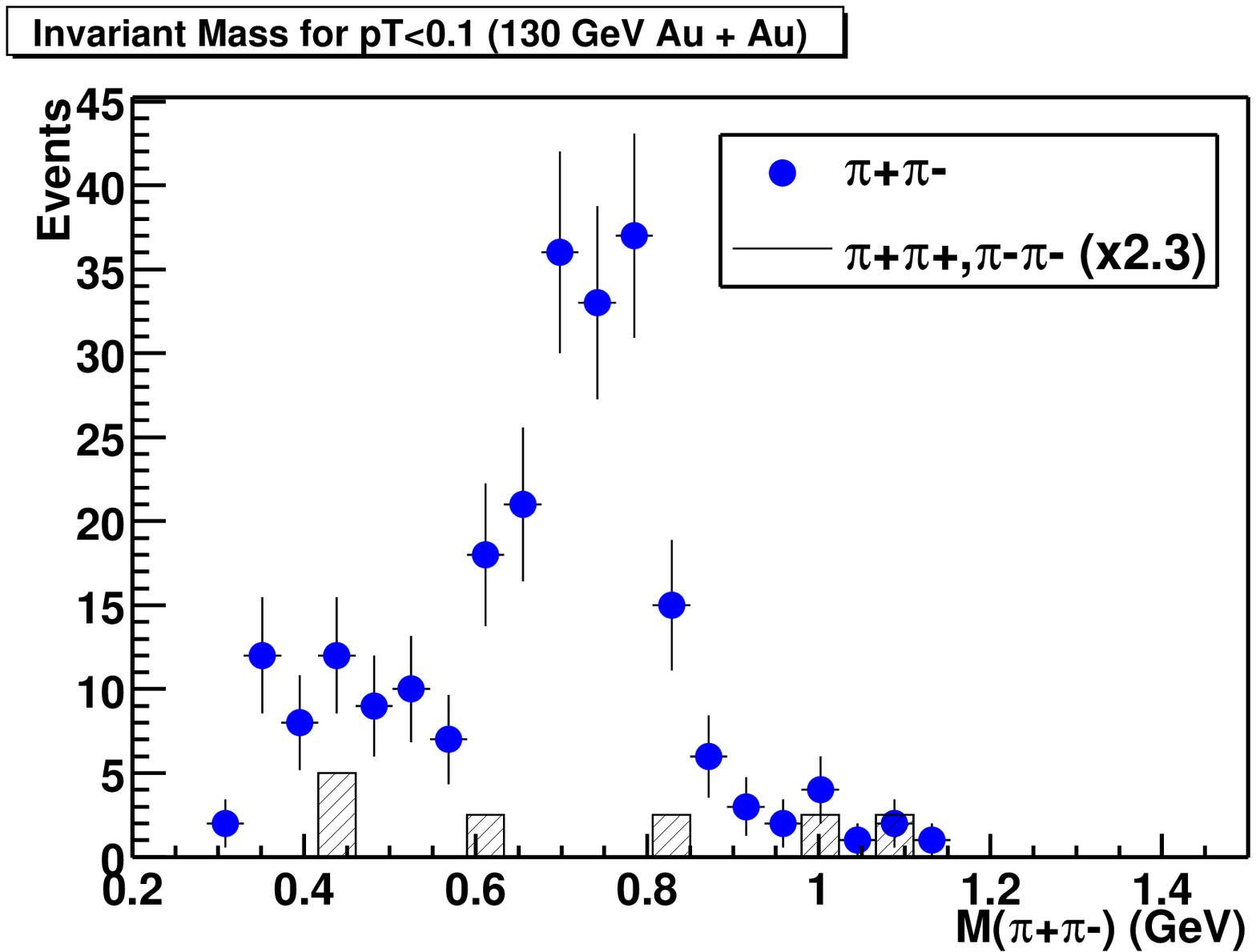}}
\vspace*{-0.2cm}
\caption{(a) The $p_T$ spectrum   2-track
events from minimum bias data.  (b) The $m_{\pi\pi}$ spectrum of 2-track events with $p_T <
100$ MeV/c.  Points are oppositely charged pairs, while the
histograms are the like-sign background.}
\vspace*{-0.3cm}
\label{fig:mbtrig}
\end{figure}

%%======== Analysis rho ====== 
Two data sets were used for the analysis of ultra-peripheral events. 
First the 'minimum bias' data set,  where  the coincident 
detection of neutrons from nuclear break up in the ZDCs is used as a trigger.
About 800k events were recorded. 
For the process $Au Au \!\rightarrow\! Au^\star Au^\star \rho^0$, i.e. $\rho^0$ production
accompanied by mutual nuclear excitation,
events were selected with exactly two tracks of opposite charge forming an common vertex
in the  interaction region. The tracks are required to have a 3-dimensional opening angle
smaller than 3 radians, i.e. they must not be perfectly back-to-back, 
to reject background from cosmic rays.
Fig.~\ref{fig:mbtrig} shows the transverse momentum 
spectrum for the selected events (points).
A clear peak, the signature for coherent coupling,  can be observed at 
$p_T\!<\!100$~MeV. Those events are compatible with coherently produced $\rho^0$ candidates.
A background model from like-sign combination pairs (shaded histogram), 
which is normalized to the signal at  $ p_T \!>\!$ 250 MeV, 
does not show such a peak.  In the invariant mass spectrum of Fig.~\ref{fig:mbtrig}b 
a clear peak is shown at the $\rho^0$ mass for the signal events at $p_T<100$~MeV.    
A clear single-neutron peak was observed for those signal events in the ZDCs. 
% ===  background 
The data contains little background: beam-gas and incoherent
photonuclear events are unlikely to deposit energy in both
ZDCs; background from grazing nuclear collisions does  not 
show a peak at low $p_T$. Nevertheless, background
from mis-identified coherently produced  $e^+e^-$ pairs, discussed below,  
contributes to the invariant mass region around $m_(\pi\pi) \!\sim\! 0.4$~GeV.
This background can be statistically subtracted.

%%% ====  Topology trigger 
For the analysis of the process $AuAu \rightarrow AuAu \rho^0$ 
a second data set was collected using a low-multiplicity topology trigger
which did not require a ZDC signal.
In the level 0 trigger, the CTB was divided into  16 coarse pixels. For a two track topology, hits were 
required in opposite pixels, while pixels in the top and the bottom acted as a veto 
to supress cosmic rays.  A fast online reconstruction - the level 3 trigger -
further removed background. With this trigger, the STAR collaboration collected about 30k 
events in 7 hours. The $\rho^0$ candidates from this data set have a transverse momentum and an 
invariant mass distributions similar to the ones already shown in Fig.~\ref{fig:mbtrig}: 
a peak at low $P_T\!<\!100$~Mev and a  peak of about 300 events around the rho mass. 
In contrast to the minimum bias data, the topology triggered data had almost
no energy deposition in the ZDC consistent with the two gold nuclei remaining
in their ground state.

\begin{figure}[!h]
\vspace*{-0.3cm}
\resizebox{.45\textwidth}{!}
{\includegraphics[height=.5\textheight,angle=270]{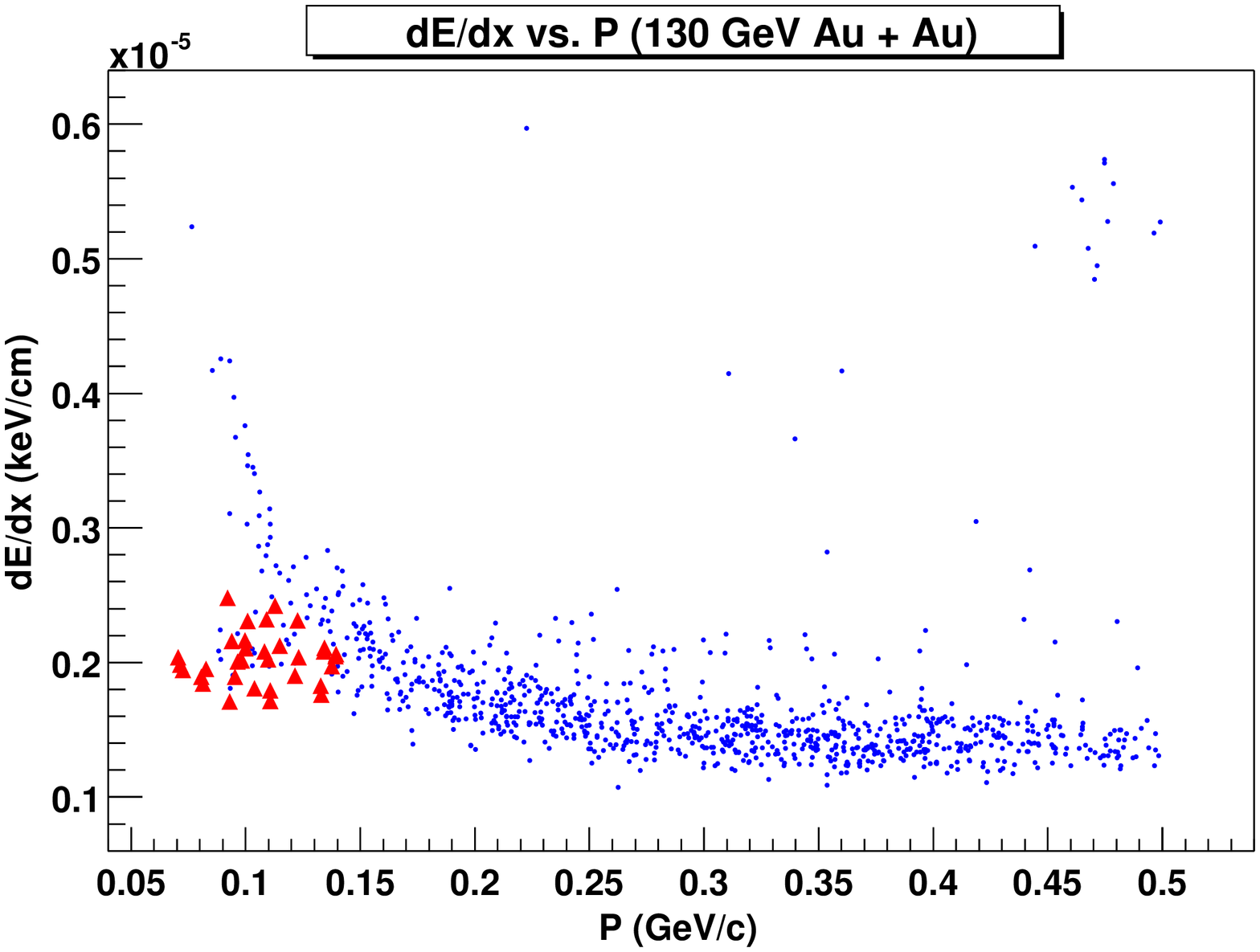}}
\resizebox{.45\textwidth}{!}
{\includegraphics[height=.5\textheight,angle=270]{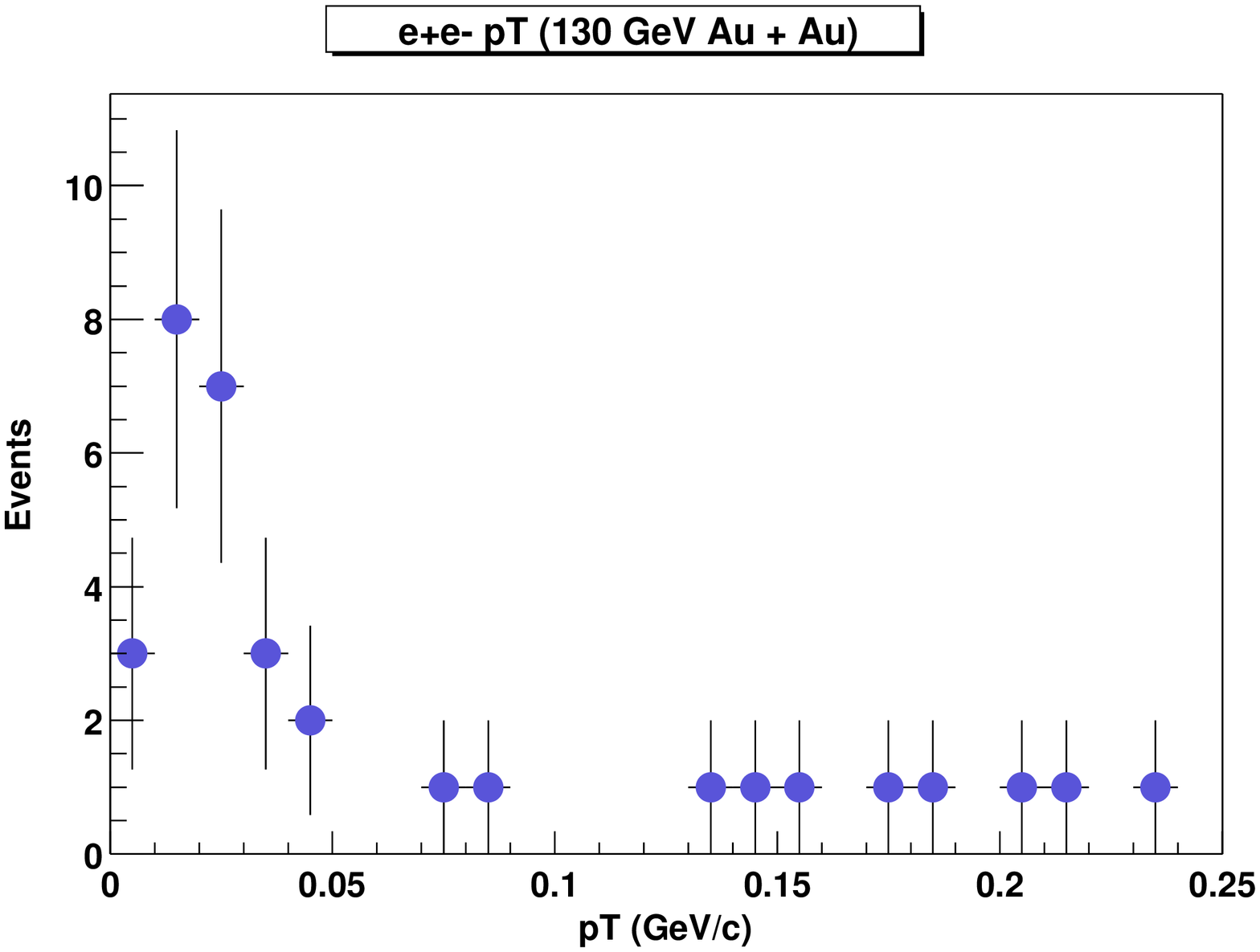}}
\caption{(a) Energy loss $dE/dx$ of tracks in the 2-track, minimum bias data;
triangles indicate events where both particles are identified as electrons. 
(b) The $p_T$ spectrum for  identified $e^+e^- pairs$. \label{fig:electrons} }
\vspace*{-0.5cm}
\end{figure}

%%%%% \section{Electron-Positron pair production}
Two-photon interactions include the purely electromagnetic process of electron-positron 
pair production as well as  single and multiple meson production.  The coupling $Z\alpha$
($0.6$ for $Au$) is large, hence  $e^+e^-$ pair production is an important
probe of quantum electrodynamics in strong fields~\cite{baurrev}.
At momenta below $140$ MeV, $e^+e^-$ pairs are identified by  
their energy loss in the TPC as shown for the minimum bias data sample
in Fig.~\ref{fig:electrons}a. Fig.~\ref{fig:electrons}b shows the $p_T$
spectrum for  identified $e^+e^-$ pairs;  a clear peak at $p_T \!< \!50$ MeV/c
identifies the process $AuAu \rightarrow Au^\star Au^\star e^+e^-$. 

%% Summary
In summary,  for the first time, 
exlcusive $\rho^0$ production  $AuAu \!\rightarrow\! Au Au \rho^0$ and  $\rho^0$
production accompanied by nuclear breakup  $Au Au \!\rightarrow \!Au^*  Au^*  \rho^0$  were 
observed in ultra-peripheral heavy  ion collisions. 
The $\rho^0$ are produced at small perpendicular momentum,
showing their coherent coupling to both nuclei. 
In addition, the coherent electromagnetic process 
$Au Au \rightarrow Au^\star Au^\star e^+ e^-$ was observed.
In 2001, RHIC will collide gold nuclei at $\sqrt{s_{NN}}= 200 $ GeV,
attempting  to reach full design luminosity. 
 Together with new  trigger algorithms, 
this will allow us to collect several orders of
magnitude larger statistics than presently available, thus 
greatly expanding the physics reach of
the STAR ultra-peripheral collisions program.  

\end{document}